\documentclass[11pt,a4paper]{article}
\usepackage[T1]{fontenc}
\usepackage[utf8]{inputenc}
\usepackage[margin=25mm]{geometry}
\usepackage{amsmath,amssymb,amsthm}
\usepackage{newtxtext,newtxmath}
\usepackage{graphicx,booktabs,array,tabularx,multirow,adjustbox,float}
\usepackage[super,sort&compress]{natbib}
\usepackage{caption,microtype,etoolbox}
\usepackage[bottom]{footmisc}
\usepackage[hidelinks,hyperfootnotes=false]{hyperref}
\graphicspath{{figures/}}
\newtheorem{proposition}{Proposition}
\AtBeginEnvironment{table}{\small\renewcommand{\arraystretch}{1.0}}
\title{Likelihood inference for cure models: Theory, computation and application}
\author{Xinyi Wang$^{1}$, Kai Liu$^{4,*}$, N. Balakrishnan$^{1}$, Wei Xu$^{2,3,*}$}
\date{}
\hypersetup{pdfauthor={Xinyi Wang; Kai Liu; N. Balakrishnan; Wei Xu},pdftitle={Likelihood inference for cure models: Theory, computation and application}}
\begin{document}
\maketitle
\begingroup
\renewcommand{\thefootnote}{}
\footnotetext{\noindent $^{1}$Department of Mathematics and Statistics, McMaster University, Hamilton, ON, Canada.\\
$^{2}$Department of Biostatistics, Princess Margaret Cancer Centre, University Health Network, Toronto, ON, Canada.\\
$^{3}$Dalla Lana School of Public Health, University of Toronto, Toronto, ON, Canada.\\
$^{4}$School of Statistics and Mathematics, Shanghai Lixin University of Accounting and Finance.\\[3pt]
$^{*}$Corresponding authors: Wei Xu and Kai Liu.\\
Email: \href{mailto:wei.xu@uhnresearch.ca}{\nolinkurl{wei.xu@uhnresearch.ca}} (Wei Xu); \href{mailto:liuk0809@lixin.edu.cn}{\nolinkurl{liuk0809@lixin.edu.cn}} (Kai Liu).}
\endgroup
\begin{abstract}
Cure models distinguish long-term freedom from an event from delayed event occurrence, but flexible lifetime distributions can complicate likelihood inference through parameter-dependent support. We develop a unified likelihood framework for power series cure models with generalized extreme value baseline log-lifetimes under independent right censoring. The framework accommodates Bernoulli, Poisson, geometric and shifted logarithmic latent-cause distributions, with covariates acting on incidence and latent lifetimes. We derive conditional augmentation laws for stochastic expectation--maximization and an explicit status-only EM formulation. We establish sufficient conditions for an unbounded unrestricted likelihood, characterize endpoint divergence in the presence of covariates, and prove existence of a finite maximum on a compact parameter set with an event-support margin. A simulation study investigates estimation and model discrimination across the four latent-cause distributions, contrasting baseline regimes, and varying censoring and cure-rate settings. An analysis of smoking cessation data compares the four models and examines covariate effects on relapse-free survival and long-term abstinence. The results clarify how lifetime support and optimization constraints shape likelihood-based inference for cure models.
\end{abstract}

\section{Introduction}
In a smoking cessation study, remaining abstinent until the end of follow-up can reflect either sustained cessation or a relapse that has not yet occurred. These possibilities lead to different questions: which characteristics are associated with long-term abstinence, and how do they affect the timing of relapse? Cure rate models address this distinction by allowing a positive probability of never experiencing the event, rather than requiring the population survival function to approach zero \cite{Amico2018}. Early cancer-survival analyses already distinguished cure from susceptible survival \cite{Boag_49,Berkson_52}. In this setting, ``cure'' is a statistical description of long-term freedom from relapse, not a directly observed clinical status. A participant with a censored relapse time cannot simply be classified as cured. The adequacy of follow-up is consequently an important part of assessing whether a cure interpretation is supported by the data \cite{Selukar2023}. Tests of practically sufficient follow-up offer a complementary assessment based on whether appreciable susceptible event risk remains beyond the observation window \cite{YuenMusta2026}.

A useful construction represents the event time as the first activation among an unobserved number of latent causes, with no event when that number is zero. Bernoulli and Poisson cause counts yield the mixture and promotion time cure models, respectively \cite{Berkson_52,chen_1999}. Power series distributions provide a common representation that also includes geometric and shifted logarithmic cause counts \cite{Noack_ps50,Cancho2013,Bao_20}. Cancho et al.\ developed likelihood inference and influence diagnostics for power series cure models \cite{Cancho2013}. These latent causes are mechanisms for the same event, rather than observed event types in a conventional competing-risks analysis. The activation rule is itself a modeling choice: recent adaptive-activation models extend beyond a fixed first-activation scheme \cite{Jiang2024}. The present analysis retains first activation and focuses on lifetime specification and likelihood computation within that scheme. The relationship between mixture and bounded-cumulative-hazard constructions provides a broader context for this modeling choice \cite{Tsodikov2003}.

The lifetime distribution is important because its tail determines how late events are distinguished from a cure component. Semiparametric mixture cure models provide an alternative in which incidence and latency have separate regression structures and the susceptible baseline distribution is left unspecified \cite{SyTaylor2000,PengDear2000}. Recent work on joint cure models with longitudinal biomarkers illustrates how assumptions about hazard functions and covariate effects can influence cure estimation \cite{Xie2025}. Here we retain proportional hazards at the latent-cause level and model baseline log-lifetimes using the generalized extreme value (GEV) distribution. Its shape parameter determines tail behavior and, for nonzero shape, a finite endpoint on one side of the log-time support \cite{coles_2001}. This provides flexibility in the baseline distribution, but does not remove the assumed proportional hazards structure. GEV-based cure models have been studied previously \cite{Li_gev_2016}, including a power series formulation for interval-censored data using Bayesian inference and Markov chain Monte Carlo methods \cite{Bao_20}. Our focus is a likelihood-based treatment of independent right-censored observations.

Censoring leaves susceptibility unobserved and makes data augmentation a natural approach to estimation within the general EM framework for incomplete-data likelihoods \cite{Dempster1977}. Balakrishnan and Pal developed EM inference for flexible cure models with Weibull lifetimes \cite{bala_16}. The stochastic expectation--maximization (SEM) algorithm samples missing quantities conditionally rather than replacing them by conditional expectations \cite{diebolt_sem_85}. Monte Carlo EM instead approximates the E-step expectation using simulated latent data \cite{WeiTanner1990}. Davies et al.\ developed SEM for Bernoulli and Poisson cure models with generalized exponential lifetimes \cite{sem_dk_2020}; Treszoks and Pal subsequently compared EM and SEM for an interval-censored destructive negative binomial cure model \cite{Treszoks2023}. These studies motivate examining stochastic augmentation in other cure-model settings, but their performance comparisons do not establish superiority for the present model. Indeed, the status-only augmentation considered here has a closed-form E-step and supplies an explicit EM comparator. SEM is therefore an alternative computational formulation whose benefits require model-specific assessment.

Flexible lifetime models also require attention to the existence of likelihood estimates. Smith established how unknown support endpoints affect likelihood existence and asymptotic behavior \cite{Smith1985}. In the GEV family, the support depends on the parameters, and standard likelihood regularity cannot be assumed without checking its conditions \cite{BucherSegers2017}. A numerical optimizer returning a finite value is not sufficient to establish a global maximum. For the present cure models, we give an explicit endpoint sequence along which the unrestricted likelihood diverges under stated conditions, and sufficient conditions for a finite maximum on a constrained parameter set. These results distinguish mathematical properties of the estimation problem from the behavior of any particular SEM implementation.

The contribution of this paper is a unified likelihood and augmentation formulation for four power series cure models with baseline log-GEV lifetimes under right censoring. We derive the conditional laws needed for susceptibility and event-time imputation, distinguish status-only from status-and-time augmentation, and provide likelihood derivatives and endpoint results that support implementation checks. The simulation study examines estimation and model discrimination across the four models, while the smoking cessation analysis examines long-term abstinence and relapse timing. This version presents selected simulation settings and numerical summaries from these analyses. The model family and SEM principle build on existing work; the contribution lies in their formulation and analysis in this setting, rather than in claiming a new latent-cause family or an established computational advantage.

Section~\ref{ch2:curemodel} specifies the models. Sections~\ref{section4_data} and \ref{ch2:sem} develop likelihood inference and SEM estimation. Section~\ref{ch2:sim} describes the simulation study and evaluation criteria; Section~\ref{ch2:realdata} presents the smoking cessation analysis. Supplementary Appendix~\ref{app_ch2} contains the derivations and proofs.

\section{Model specification}
\label{ch2:curemodel}

For individual $i$, let $M_i$ denote the unobserved number of latent causes. Conditional on $M_i = m$ (with $m=0,1,2,\dots$), let $W_{ik^\star}$, for $k^\star=1,2,\dots,m$, denote the independent, unobserved times to the event associated with those causes. These times have the common survival function $S(t) = 1 - F(t)$ and density $f(t) = -S'(t)$. Under the first activation scheme, the event occurs when the first latent cause is activated. For $M_{i} \geq 1$, the event time is $\:Y_{i}= \text{min} \{W_{i1}, W_{i2}, \dots, W_{iM_{i}} \}$. When $M_i = 0$, the event never occurs, so we set $Y_i = \infty$ and have $P(Y_i=\infty|M_i=0)=1$.

The number of causes $M_i$ follows a power series distribution of the form
\begin{equation}
P(M_i = m) = \frac{a_m \theta_i^m}{\sum_{m=0}^\infty a_m \theta_i^m} = \frac{a_m \theta_i^m}{G(\theta_i)}, \qquad m = 0,1,2,\dots,
\label{powerseries}
\end{equation}
Here, $a_m \geq 0$, $a_0>0$, and at least one $a_m$ with $m\geq1$ is positive, and $\theta_i \in (0,s)$ ensures that $G(\theta_i)=\sum_{m=0}^\infty a_m \theta_i^m < \infty$. We assume that $G$ is at least three times differentiable. Table~\ref{tab:cure_models} specifies the four latent-cause distributions used in this paper.

Let $C_i$ denote the censoring time. Conditional on the covariates, censoring is independent of the latent event process, and its distribution has parameters distinct from those of the event model. The observed time is $T_i = \min\{Y_i, C_i\}$.

Averaging over the distribution of $M_i$ gives the population survival function of $Y_i$:
\begin{align}
	S_p(t_{i}) &= P(Y_{i} > t_{i}) = \sum_{m = 0}^{\infty} S(t_{i})^m \frac{a_m \theta_{i}^m}{G(\theta_{i})} = \frac{ G(\theta_i S(t_i))}{G(\theta_i)}.
    \end{align}
    Differentiating this expression gives the population density function:
    \begin{align}
	f_p (t_{i}) & = \frac{G'(\theta_{i}S(t_{i})) \theta_{i} f(t_{i})}{G(\theta_{i})},
    \end{align}
where $G'(x) = d G(x)/ dx$. The corresponding population hazard function is
    \begin{align}
	\lambda_p (t_{i}) & = \frac{G'(\theta_{i}S(t_{i}))\theta_{i}f(t_{i})}{G(\theta_{i}S(t_{i})) }.
    \end{align}
The cure fraction, defined as the probability that an individual never experiences the event of interest, is
    \begin{align}
	p_{0i} &= \lim_{t \rightarrow \infty} S_{p,i}(t) =\frac{G(0)}{G(\theta_i)} = \frac{a_0}{G(\theta_i)},
\end{align}
since $\lim_{t\to\infty} S(t)=0$ and $G(0)=a_0$. 
Table~\ref{tab:cure_models} summarizes the cure rate models obtained when the number of latent causes follows a Bernoulli, Poisson, geometric, or logarithmic distribution. The table collects the power series coefficients, generating functions, parameter domains, cure fractions and incidence links. The population survival and density for all four models follow from the unified formulas above. We assume proportional hazards for $S(t_i)$, with $S(t_i) = S_0(t_i)^{\text{exp}(\boldsymbol{z}_{i}^T \boldsymbol{\beta})}$ and $\lambda = \lambda_0(t_i)\text{exp}(\boldsymbol{z}_{i}^T \boldsymbol{\beta})$. The baseline survival and hazard functions are derived from the GEV distribution. Throughout, $S(t_i)$ and $f(t_i)$ suppress individual covariates; more explicitly, with $a_i=\exp(\boldsymbol z_i^T\boldsymbol\beta)$,
\begin{equation}\label{eq:latent_ph}
 S_i(t)=S_0(t)^{a_i},\qquad f_i(t)=a_i f_0(t)S_0(t)^{a_i-1}=a_i\lambda_0(t)S_i(t).
\end{equation}
This proportional hazards assumption concerns each latent cause, not necessarily the population or the event time conditional on susceptibility. The latter has survival $(S_{p,i}(t)-p_{0i})/(1-p_{0i})$.
The logarithmic model in this paper is the shifted logarithmic-series distribution on $\{0,1,\ldots\}$: $M_i+1$ has the ordinary logarithmic-series distribution. For the Bernoulli model, $a_0=a_1=1$ and $a_m=0$ for $m>1$. The vector $\boldsymbol{\beta}$ contains the regression coefficients for the covariates $\boldsymbol{z}_i$.

\begin{table}[tbp]
\centering
\caption{Power series specifications, cure fractions and incidence links for the four cure models.}
\label{tab:cure_models}
\renewcommand{\arraystretch}{1.6}
\begin{tabular*}{\textwidth}{@{\extracolsep{\fill}}lccccc@{}}
\toprule
Model & $a_m$ & $G(\theta_i)$ & Domain & $p_{0i}$ & $\theta_i(\eta_i)$ \\
\midrule
Bernoulli & $\mathbf1\{m=0,1\}$ & $1+\theta_i$ & $\theta_i>0$ & $\dfrac{1}{1+\theta_i}$ & $e^{\eta_i}$ \\
Poisson & $1/m!$ & $e^{\theta_i}$ & $\theta_i>0$ & $e^{-\theta_i}$ & $e^{\eta_i}$ \\
Geometric & $1$ & $\dfrac{1}{1-\theta_i}$ & $0<\theta_i<1$ & $1-\theta_i$ & $\dfrac{e^{\eta_i}}{1+e^{\eta_i}}$ \\
Shifted logarithmic & $\dfrac{1}{m+1}$ & $\dfrac{-\log(1-\theta_i)}{\theta_i}$ & $0<\theta_i<1$ & $\dfrac{-\theta_i}{\log(1-\theta_i)}$ & $\dfrac{e^{\eta_i}}{1+e^{\eta_i}}$ \\
\bottomrule
\end{tabular*}
\par\smallskip
\begin{minipage}{\textwidth}\footnotesize
Here $m=0,1,\ldots$ and $\eta_i=\boldsymbol x_i^T\boldsymbol b$. The Poisson specification gives the promotion time cure model. All four specifications have $a_0=1$.
\end{minipage}
\end{table}

We model a baseline latent lifetime $W_0$ through $\log W_0\sim\mathrm{GEV}(\mu,\sigma,\gamma)$, rather than assigning a GEV law to the observed censored time $T_i$. Write $z(t)=(\log t-\mu)/\sigma$ and $h(t)=1+\gamma z(t)$, where $t>0$, $\sigma>0$ and $\mu,\gamma\in\mathbb R$. On the interior of the support,
\begin{equation}\label{eq:ch1_base_surv}
 S_0(t)=\begin{cases}
 1-\exp\{-h(t)^{-1/\gamma}\},&\gamma\ne0,\ h(t)>0,\\
 1-\exp\{-\exp[-z(t)]\},&\gamma=0.
 \end{cases}
\end{equation}
For $\gamma>0$, set $S_0(t)=1$ when $t\leq \exp(\mu-\sigma/\gamma)$; for $\gamma<0$, set $S_0(t)=0$ when $t\geq \exp(\mu-\sigma/\gamma)$. Also $S_0(t)=1$ for $t\leq0$. These extensions specify the distribution without ambiguous powers of zero. On the support interior,
\begin{equation}\label{eq:baseline_density}
 f_0(t)=\begin{cases}
 \dfrac{h(t)^{-1/\gamma-1}\exp\{-h(t)^{-1/\gamma}\}}{\sigma t},&\gamma\ne0,\\[.5em]
 \dfrac{e^{-z(t)}\exp\{-e^{-z(t)}\}}{\sigma t},&\gamma=0,
 \end{cases}
\end{equation}
and $f_0=0$ outside the support (endpoint values are immaterial for a continuous density). Consequently,
\begin{equation}\label{eq:ch1_base_hazard}
 \lambda_0(t)=\frac{f_0(t)}{S_0(t)}=\begin{cases}
 \dfrac{h(t)^{-1/\gamma-1}}{\sigma t\,[\exp\{h(t)^{-1/\gamma}\}-1]},&\gamma\ne0,\\[.5em]
 \dfrac{e^{-z(t)}}{\sigma t\,[\exp\{e^{-z(t)}\}-1]},&\gamma=0.
 \end{cases}
\end{equation}
The hazard is zero below a positive lower endpoint and is undefined where $S_0=0$. The formulas agree with the standard GEV parameterization \cite{BucherSegers2017}. The Gumbel, Fr\'echet and reverse-Weibull classifications apply to the baseline \emph{log-time}, with $\gamma=0$, $\gamma>0$ and $\gamma<0$, respectively \cite{coles_2001}. Exponentiating log-time changes tail properties; these labels should not be read as ordinary Weibull or Fr\'echet laws for time itself. Supplementary Appendices~\ref{app:part1}--\ref{app:part4} give the derivatives, limits and inversion formulas.

\section{Data and likelihood}
\label{section4_data}
Let $\mathcal O=\{(t_i,\delta_i,\boldsymbol x_i,\boldsymbol z_i):i=1,\ldots,n\}$, where $T_i=\min(Y_i,C_i)$ and $\delta_i=\mathbf1\{Y_i\leq C_i\}$. Individuals are assumed independent conditional on their covariates. Let $J_i=\mathbf1\{M_i\geq1\}$ denote susceptibility, and write $\boldsymbol\xi=(\boldsymbol b^T,\boldsymbol\beta^T,\mu,\sigma,\gamma)^T$. Suppressing parameter arguments, the observed likelihood and log-likelihood are
\begin{align}
 L_o(\boldsymbol\xi;\mathcal O)&\propto\prod_{i=1}^n f_{p,i}(t_i)^{\delta_i}S_{p,i}(t_i)^{1-\delta_i},\label{non-infoL}\\
 \ell_o(\boldsymbol\xi)&=\sum_i\{\delta_i\log f_{p,i}(t_i)+(1-\delta_i)\log S_{p,i}(t_i)\}.\label{eq:observed_loglik}
\end{align}
Terms involving only the censoring distribution have been omitted.

\paragraph{Augmenting cure status only.}
Put $\mathcal E=\{i:\delta_i=1\}$ and $\mathcal C=\{i:\delta_i=0\}$. For $i\in\mathcal E$, $J_i=1$ is known. If only the missing $J_i$ are augmented, the log-likelihood is
\begin{equation}\label{eq:status_loglik}
\begin{split}
 \ell_J(\boldsymbol\xi)&=\sum_{i\in\mathcal E}\log f_{p,i}(t_i)\\
 &\quad+\sum_{i\in\mathcal C}\{(1-J_i)\log p_{0i}+J_i\log[S_{p,i}(t_i)-p_{0i}]\}.
\end{split}
\end{equation}
The four explicit substitutions are given in Supplementary Appendix~\ref{app:part3}. Equation~\eqref{eq:status_loglik} augments cure status only; it does not impute event times.

\paragraph{Augmenting both status and event time.}
If a censored susceptible individual's event time $y_i^*>t_i$ is also sampled, set $\widetilde y_i=t_i$ for observed events and $\widetilde y_i=y_i^*$ for imputed susceptible events. The appropriate joint log-likelihood is instead
\begin{equation}\label{eq:full_aug_loglik}
 \ell_A(\boldsymbol\xi)=\sum_{i:J_i=1}\log f_{p,i}(\widetilde y_i)
                  +\sum_{i:J_i=0}\log p_{0i}.
\end{equation}
The factor $f_{p,i}$ is the joint density of a finite event and $J_i=1$; it must not be divided by $1-p_{0i}$ unless the corresponding susceptibility probability is also included. Equations~\eqref{eq:status_loglik} and \eqref{eq:full_aug_loglik} are distinct valid augmentations, not interchangeable M-step objectives.

\section{Stochastic expectation--maximization algorithm}
\label{ch2:sem}
The following formulation makes explicit the status-and-time augmentation described in the original algorithm \cite{diebolt_sem_85}. A status-only alternative is given in Supplementary Appendix~\ref{app:part5}.

\paragraph{Step 1: Initialization.}
The reported initialization evaluates $\ell_o$ at 2000 candidate parameter vectors and selects the best feasible candidate $\boldsymbol\xi^{(0)}$. Candidate generation and parameter bounds must be specified for reproducibility. A feasible vector has $\sigma>0$, admissible $\theta_i$, and positive event densities at all observed event times.

\paragraph{Step 2: Sample susceptibility.}
At iteration $r$, hold $\boldsymbol\xi^{(r)}$ fixed. For each $i\in\mathcal C$, independently sample $J_i^{(r+1)}\sim\mathrm{Bernoulli}(q_i^{(r)})$, where
\begin{equation}\label{eq:sem_step2}
 q_i^{(r)}=P(J_i=1\mid Y_i>t_i;\boldsymbol\xi^{(r)})
 =\frac{S_{p,i}(t_i;\boldsymbol\xi^{(r)})-p_{0i}(\boldsymbol\xi^{(r)})}{S_{p,i}(t_i;\boldsymbol\xi^{(r)})}.
\end{equation}
For $i\in\mathcal E$, set $J_i^{(r+1)}=1$. The symbol $q_i^{(r)}$ is distinct from the unconditional cure probability $p_{0i}$.

\paragraph{Step 3: Sample censored susceptible event times.}
Conditional on $J_i^{(r+1)}=1$ and $Y_i>t_i$, the proper conditional density and distribution are, for $y>t_i$,
\begin{align}
 g_i(y)&=\frac{f_{p,i}(y)}{S_{p,i}(t_i)-p_{0i}},\label{eq:ch1:sem_step3}\\
 H_i(y)&=\frac{S_{p,i}(t_i)-S_{p,i}(y)}{S_{p,i}(t_i)-p_{0i}},\label{eq:conditional_cdf}
\end{align}
with all terms evaluated at $\boldsymbol\xi^{(r)}$. Draw $U_i\sim\mathrm{Uniform}(0,1)$ and solve
\begin{equation}\label{eq:conditional_inverse}
 S_{p,i}(y_i^*)=p_{0i}+(1-U_i)[S_{p,i}(t_i)-p_{0i}].
\end{equation}
When $q_i^{(r)}=0$ no finite event is sampled. For $J_i^{(r+1)}=0$, set $Y_i=\infty$ and use the cure mass, not a density at infinity. The earlier defective CDF divided by $S_{p,i}(t_i)$ gives the same sampler if its uniform upper bound is exactly $q_i^{(r)}$; Supplementary Appendix~\ref{app:part4} proves this equivalence.

\paragraph{Step 4: Maximization.}
Using the sampled augmentation, seek a feasible maximizer
\begin{equation}\label{eq:mstep}
 \boldsymbol\xi^{(r+1)}\in\operatorname*{arg\,max}_{\boldsymbol\xi\in\Theta_A}
       \ell_A(\boldsymbol\xi;\boldsymbol J^{(r+1)},\widetilde{\boldsymbol y}^{(r+1)}).
\end{equation}
Here $\Theta_A$ requires positive density at every observed or imputed susceptible event. Numerical optimization is implemented using the L-BFGS-B method in the R function \texttt{optim}, with a lower bound of $10^{-6}$ on $\sigma$ and no finite box bounds on the regression coefficients, $\mu$ or $\gamma$. Box constraints alone do not enforce the GEV support restrictions, which depend jointly on $(\mu,\sigma,\gamma)$ and event times. A numerical local optimum is not a proof that the global argmax exists or has been found. Proposition~\ref{prop:unbounded} in Supplementary Appendix~\ref{app:part7} proves unboundedness under its stated parameter-space conditions; the full proof and divergence rate \eqref{eq:loglik_divergence_rate} are given there. Proposition~\ref{prop:endpoint_covariates} in the same appendix covers fixed covariate effects. Proposition~\ref{prop:compact} in Supplementary Appendix~\ref{app:compact} states and proves sufficient conditions for a finite constrained maximum.

\paragraph{Steps 5--6: Iteration and averaging.}
Repeat Steps 2--4 for $R$ iterations. For burn-in $r^*$ and retained indices $\mathcal K=\{r^*+3,r^*+6,\ldots\}\cap\{1,\ldots,R\}$, define
\begin{equation}\label{eq:sem_average}
 \widehat{\boldsymbol\xi}_{\rm SEM}=\frac1{|\mathcal K|}\sum_{r\in\mathcal K}\boldsymbol\xi^{(r)}.
\end{equation}
This specifies averaging on the original parameter scale. Averaging transformed parameters defines a different estimator. The resulting vector must itself be checked for feasibility because the support-constrained parameter set need not be convex. An inner optimizer tolerance is distinct from convergence of the stochastic outer iterations. SEM does not generally increase $\ell_o$ at each iteration; stationary behavior and negligible Monte Carlo error require assessment rather than a deterministic successive-iterate stopping rule. Neither an exact finite-sample MLE identity for the average nor an ergodicity theorem for this support-dependent model is asserted here.

\paragraph{Uncertainty quantification.}
For a regular, identifiable model and a feasible estimate sufficiently close to an interior observed-likelihood maximum, the observed-information approximation is
\begin{equation}\label{eq:observed_information}
 \widehat{\mathcal I}_o=-\nabla^2\ell_o(\widehat{\boldsymbol\xi}),\qquad
 \widehat V=\widehat{\mathcal I}_o^{-1},\qquad
 \widehat{\mathrm{se}}_j=\sqrt{\widehat V_{jj}}.
\end{equation}
This uses the Hessian of the \emph{total observed-data} log-likelihood; there is no additional factor $1/n$. Expected Fisher information is instead $\mathcal I(\boldsymbol\xi)=E_{\boldsymbol\xi}[-\nabla^2\ell_o(\boldsymbol\xi)]$, when the expectation and derivatives exist. An inverse completed-data Hessian is not a substitute for observed information: the missing-information contribution must also be accounted for, as in the identity of Louis \cite{Louis1982}. Under the regular approximation, a componentwise interval is
\begin{equation}\label{eq:wald}
 \widehat\xi_j\ \pm\ z_{1-\alpha/2}\sqrt{\widehat V_{jj}}.
\end{equation}
For $\sigma>0$, a delta-method interval on the log scale is
\begin{equation}\label{eq:scale_ci}
 \exp\left\{\log\widehat\sigma\ \pm\ z_{1-\alpha/2}
       \frac{\widehat{\mathrm{se}}(\widehat\sigma)}{\widehat\sigma}\right\}.
\end{equation}
These formulas require nonsingular positive information and suitable regularity; they do not account automatically for finite-run SEM error. GEV support depends on its parameters, so regularity cannot be presumed for all simulation settings. In particular, the usual uncensored GEV normal theory does not cover $\gamma\leq-1/2$; its conditions cannot simply be transferred to this cure model either \cite{BucherSegers2017}.

\subsection{SEM parameter trajectories}
Trace plots can reveal drift or instability but do not prove stationarity or accurate likelihood maximization. Multiple starts, retained-run variability and observed-likelihood comparisons should accompany the plotted trajectories \cite{Nielsen_00}.

\section{Simulation study}
\label{ch2:sim}
The simulation study examines estimation and model discrimination across Bernoulli, Poisson, geometric and shifted logarithmic latent-cause distributions, Type I, II and III GEV baselines, and varying censoring and cure-rate settings. All four candidate models are fitted to the simulated data. The three parameter vectors below are selected examples from the eleven settings in the study, rather than an exhaustive account of its scope. We consider sample sizes $n=500$ and $n=1000$, with a binary covariate $X_i\sim\mathrm{Bernoulli}(c)$, where $0<c<1$, entering both incidence and latency. The incidence predictor is $b_0+b_1X_i$, and the latent-time proportional hazards multiplier is $\exp(\beta_1X_i)$. The incidence predictors use the model-specific links in Table~\ref{tab:cure_models}.

A Type I example uses $b_0=-0.7$, $b_1=0.52$, $\beta_1=0.64$, $\mu=0.2$, $\sigma=0.17$ and $\gamma=0$. A Type II example uses $b_0=-0.6$, $b_1=0.15$, $\beta_1=0.3$, $\mu=0.1$, $\sigma=0.15$ and $\gamma=1.6$. A Type III example uses $b_0=-0.89$, $b_1=0.53$, $\beta_1=0.875$, $\mu=0.14$, $\sigma=0.44$ and $\gamma=-1.1$.

Under the model, event times can be generated using the population inversion in Supplementary Appendix~\ref{app:part4}, including the cure mass at infinity. For independent censoring $C_i\sim\mathrm{Exp}(cc)$, with $cc$ a rate, the observations are $T_i=\min(Y_i,C_i)$ and $\delta_i=\mathbf1\{Y_i\le C_i\}$. The generating and fitted models are distinguished when assessing estimation and model discrimination. The censoring rate and cure-model parameters vary across settings; the displayed parameter vectors are selected to describe the three baseline types.

The Type III design, with $\gamma=-1.1$, is particularly relevant to the endpoint behavior studied in Supplementary Appendix~\ref{app:part7}. Numerical assessment must state the optimization domain, including any bounds or endpoint margin; a finite optimizer output cannot be interpreted as an unrestricted global maximum. The existence result on a compact set does not justify Wald inference in this regime.

For $N_{\rm MC}$ independently generated datasets and a common scalar estimand $\psi$, define
\begin{align}
 \widehat{\mathrm{Bias}}&=\frac1{N_{\rm MC}}\sum_{r=1}^{N_{\rm MC}}(\widehat\psi_r-\psi),\\
 \widehat{\mathrm{RMSE}}&=\left\{\frac1{N_{\rm MC}}\sum_{r=1}^{N_{\rm MC}}(\widehat\psi_r-\psi)^2\right\}^{1/2},\\
 \widehat{\mathrm{CP}}&=\frac1{N_{\rm MC}}\sum_{r=1}^{N_{\rm MC}}\mathbf1\{\psi\in\mathrm{CI}_r\},\qquad
 \widehat{\mathrm{MCSE}}(\widehat{\mathrm{CP}})=\sqrt{\frac{\widehat{\mathrm{CP}}(1-\widehat{\mathrm{CP}})}{N_{\rm MC}}}.
\end{align}
$N_{\rm MC}$ is the dataset replication count, not the number of SEM iterations. Under model misspecification, coefficients associated with different links need not represent the same estimand; common cure probabilities or survival predictions provide interpretable comparisons. The same binomial Monte Carlo error formula applies to a model-selection rate. Separating performance measures from their Monte Carlo uncertainty is a central principle of simulation-study reporting \cite{Morris2019}.

The numerical presentation in this version is selective: a Poisson simulation summary is included in Supplementary Appendix~\ref{app:poisson_example}. This choice of displayed results does not restrict the simulation study to the Poisson model or to the three parameter vectors above.

For model discrimination, AIC \cite{Akaike_74}, BIC \cite{Schwarz_78}, and the conventional AICc correction \cite{aicc} must use the observed-data likelihood. When all candidates have the same parameter count and sample size, their rankings coincide; they do not provide three independent confirmations of model preference.

\section{Analysis of smoking cessation data}

\label{ch2:realdata}

We apply the proposed cure rate models to smoking cessation data from southeastern Minnesota, United States. The study enrolled 223 smokers between 1986 and 1989 and randomly assigned them to a smoking intervention (SI) or usual care (UC). The recorded covariates included sex (136 men coded as 0 and 87 women coded as 1), years of smoking (12--46), and average daily cigarette consumption over the preceding 10 years (5--60).

The UC group included 54 participants, and the SI group included 169. By the end of the study, 65 participants had relapsed. The remaining 158 participants were right-censored, giving a censoring proportion of $158/223\approx0.709$ (70.9\%). Their cure status was unobserved.

In the analysis input, the 65 event durations were calculated as the difference between the recorded variables \texttt{Time2} and \texttt{Time1}; all 158 censored durations were assigned a value of 5. The event durations ranged from 0.770 to 4.170. There were no missing values in the four model covariates, analysis time or event indicator.

Treatment was coded as 0 for UC and 1 for SI. Both the incidence and latency predictors included sex, smoking duration, treatment and cigarette consumption, in that order, on their original scales without centering or standardization. The incidence predictor included an intercept, whereas the latency predictor did not. Thus each model contained 12 free parameters: five incidence coefficients, four latency coefficients and the three GEV parameters. The fitting routines were configured for 3000 iterations, with a maximum of 20000 iterations for each call to \texttt{optim}.

All four candidate models are fitted to the smoking cessation data. Table~\ref{tab:poisson_main} presents the Poisson parameter estimates, standard errors and 95\% confidence intervals as a selected detailed summary; Table~\ref{table:ch2:ll} reports the model-comparison results for all four candidates. The restriction of the detailed coefficient table and survival plot to Poisson is a presentation choice, not a restriction of the analysis to that model.

\begin{figure}[tbp]
\centering
\includegraphics[width=.8\linewidth]{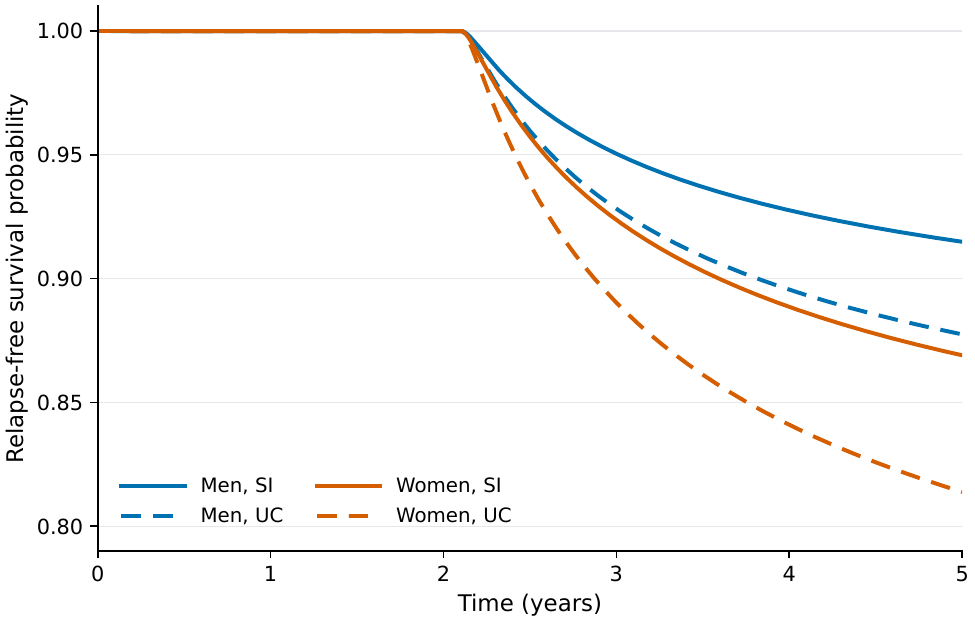}
\caption{Relapse-free survival for four sex--treatment profiles under the Poisson cure model, calculated from the parameter estimates displayed in Table~\ref{tab:poisson_main}, with 30 years of smoking and six cigarettes daily. Blue and orange denote men and women, respectively; solid and dashed lines denote smoking intervention (SI) and usual care (UC), respectively.}
\label{fig:poisson_survival}
\end{figure}

Figure~\ref{fig:poisson_survival} displays fitted relapse-free survival curves by sex and treatment for covariate profiles with 30 years of smoking and six cigarettes per day. Under the specified incidence links, a positive coefficient corresponds to a lower long-term cure probability, holding the other covariates fixed.

\begin{table}[tbp]
\centering
\caption{Parameter estimates, standard errors and 95\% confidence intervals for the Poisson cure model fitted to the smoking cessation data.}
\label{tab:poisson_main}
\renewcommand{\arraystretch}{1.12}
\begin{tabular*}{\textwidth}{@{\extracolsep{\fill}}lrrrr@{}}
\toprule
Parameter & Estimate & SE & \multicolumn{2}{c}{95\% CI} \\
\cmidrule(l){4-5}
 & & & Lower & Upper \\
\midrule
\multicolumn{5}{@{}l}{\textit{Incidence component}} \\
Intercept ($b_0$) & $-0.187$ & $0.446$ & $-1.062$ & $0.687$ \\
Sex ($b_1$) & $0.684$ & $0.202$ & $0.288$ & $1.080$ \\
Smoking duration ($b_2$) & $-0.014$ & $0.012$ & $-0.038$ & $0.011$ \\
Treatment ($b_3$) & $-0.433$ & $0.211$ & $-0.846$ & $-0.020$ \\
Cigarette consumption ($b_4$) & $0.038$ & $0.009$ & $0.021$ & $0.056$ \\
\addlinespace[3pt]
\multicolumn{5}{@{}l}{\textit{Latency component}} \\
Sex ($\beta_1$) & $-0.252$ & $0.305$ & $-0.850$ & $0.346$ \\
Smoking duration ($\beta_2$) & $-0.002$ & $0.013$ & $-0.027$ & $0.023$ \\
Treatment ($\beta_3$) & $0.054$ & $0.324$ & $-0.582$ & $0.690$ \\
Cigarette consumption ($\beta_4$) & $-0.025$ & $0.013$ & $-0.050$ & $-0.001$ \\
\addlinespace[3pt]
\multicolumn{5}{@{}l}{\textit{Baseline log-GEV distribution}} \\
Location ($\mu$) & $3.274$ & $0.659$ & $1.982$ & $4.567$ \\
Scale ($\sigma$) & $7.038$ & $2.042$ & $3.034$ & $11.041$ \\
Shape ($\gamma$) & $2.777$ & $0.231$ & $2.324$ & $3.230$ \\
\bottomrule
\end{tabular*}
\par\smallskip
\begin{minipage}{\textwidth}\footnotesize
SE: standard error; CI: confidence interval.
\end{minipage}
\end{table}

Table~\ref{table:ch2:ll} reports the negative log-likelihood, AIC, BIC, and AICc for Models 1--4. The Poisson model has the lowest reported negative log-likelihood, AIC, BIC and AICc among the four fitted candidates.

\begin{table}[tbp]
	\centering
	\caption{Negative log-likelihood, AIC, BIC, and AICc for the four models fitted to the smoking cessation data. Bold entries indicate the minimum in each column.}
	\label{table:ch2:ll}
    
	\renewcommand{\arraystretch}{1.12}
\begin{tabular*}{\textwidth}{@{\extracolsep{\fill}}lrrrr@{}}
		\toprule
		 Model & $-\ell_o$ & AIC & BIC & AICc\\
		\midrule
		 Bernoulli   & 295.368         & 614.735          & 655.621         & 616.221        \\
		 Poisson    & \textbf{292.638} & \textbf{609.277} & \textbf{650.163} & \textbf{610.762}\\
		 Geometric  & 296.250        & 616.500         & 657.386          & 617.985        \\
		 Logarithmic & 294.602         & 613.205          & 654.091          & 614.690      \\
		\bottomrule
	\end{tabular*}
    
\end{table}

\section{Concluding remarks}
\label{conclusion}

We developed likelihood and augmentation formulations for four power series cure models with a baseline log-GEV distribution. The conditional laws support both status-only EM and stochastic augmentation, but do not establish a computational advantage of SEM. The endpoint results demonstrate why unrestricted likelihood maximization and inference require care, while compactness and an event-support margin provide a sufficient constrained existence condition. The simulation study examines estimation and model discrimination across the four latent-cause distributions and contrasting baseline, censoring and cure-rate settings. The smoking cessation analysis fits and compares all four models, with the Poisson model attaining the lowest reported information criteria. Selected settings and a Poisson simulation summary are presented here, together with detailed Poisson application estimates and survival profiles; this selection condenses the numerical presentation while retaining the broader scope of the study.

The analysis uses a fully parametric cure rate model. Semiparametric proportional hazards cure models \cite{SyTaylor2000,PengDear2000} and nonparametric incidence and latency estimation, as implemented in \texttt{npcure} \cite{LopezCheda2021}, provide useful directions for sensitivity analysis of the lifetime specification. Extending the present power series construction along these lines is a direction for future work.

\clearpage
\bibliography{references}
\clearpage\appendix
\section{Supplementary material: mathematical derivations and proofs}\label{app_ch2}
\subsection{Population distribution and the cure component}\label{app:part1}
\begin{proposition}[Distribution induced by latent causes]\label{prop:population}
Suppose $M_i$ has the power series law in \eqref{powerseries}, and conditional on covariates and $M_i$, the latent times are independent with a proper continuous survival $S_i$. Then, for finite $t$,
\begin{equation}
 S_{p,i}(t)=\frac{G(\theta_i S_i(t))}{G(\theta_i)},\quad
 f_{p,i}(t)=\frac{\theta_i f_i(t)G'(\theta_i S_i(t))}{G(\theta_i)},\quad
 p_{0i}=\frac{a_0}{G(\theta_i)}.
\end{equation}
The finite-time density integrates to $1-p_{0i}$, and the remaining mass $p_{0i}$ is at infinity.
\end{proposition}
\begin{proof}
For $m\geq1$, independence gives $P(Y_i>t\mid M_i=m)=S_i(t)^m$; for $m=0$ the probability is one. Summing the nonnegative terms gives the generating-function ratio. Differentiation on the support interior, using $S_i'=-f_i$, gives the stated density and $\lambda_{p,i}=f_{p,i}/S_{p,i}$. Properness gives $S_i(0+)=1$ and $S_i(\infty)=0$, so $S_{p,i}(0+)=1$ and $S_{p,i}(\infty)=a_0/G(\theta_i)$. Integrating $-S_{p,i}'$ gives $1-p_{0i}$. Thus
\begin{equation}\label{eq:susceptible_law}
 S_{u,i}(t)=P(Y_i>t\mid J_i=1)=\frac{S_{p,i}(t)-p_{0i}}{1-p_{0i}},\qquad
 f_{u,i}(t)=\frac{f_{p,i}(t)}{1-p_{0i}}
\end{equation}
are proper susceptible distributions, and $S_{p,i}=p_{0i}+(1-p_{0i})S_{u,i}$.
\end{proof}

The required generating functions and derivatives are
\begin{align}
 G_{\rm Ber}(x)&=1+x,&G'_{\rm Ber}(x)&=1,\\
 G_{\rm Poi}(x)&=e^x,&G'_{\rm Poi}(x)&=e^x,\\
 G_{\rm Geo}(x)&=(1-x)^{-1},&G'_{\rm Geo}(x)&=(1-x)^{-2},\\
 G_{\rm Log}(x)&=\frac{-\log(1-x)}{x},&
 G'_{\rm Log}(x)&=\frac{x/(1-x)+\log(1-x)}{x^2}.
\end{align}
The last pair extends continuously to $G_{\rm Log}(0)=1$ and $G'_{\rm Log}(0)=1/2$. Indeed,
\begin{equation}
 G_{\rm Log}(x)=\sum_{m=0}^{\infty}\frac{x^m}{m+1},\qquad
 \frac{x}{1-x}+\log(1-x)=\sum_{k=2}^{\infty}\frac{k-1}{k}x^k>0\quad(0<x<1).
\end{equation}
Substituting these functions into Proposition~\ref{prop:population} proves every population survival, density and cure fraction in Table~\ref{tab:cure_models}. In particular, the logarithmic density is positive despite its original expression containing negative logarithms. The ordinary logarithmic-series law on $\{1,2,\ldots\}$ would have no atom at zero and hence no cure fraction; the shift is essential.

\subsection{Log-GEV derivatives, limits and inversion}\label{app:part2}
Let $z=(\log t-\mu)/\sigma$, $h=1+\gamma z$ and $w=h^{-1/\gamma}$. For $\gamma\ne0$ and $h>0$,
\begin{equation}
 \frac{dh}{dt}=\frac{\gamma}{\sigma t},\qquad
 \frac{dw}{dt}=-\frac{h^{-1/\gamma-1}}{\sigma t},\qquad
 -\frac{d}{dt}(1-e^{-w})=\frac{h^{-1/\gamma-1}e^{-w}}{\sigma t}.
\end{equation}
This proves \eqref{eq:baseline_density}; division by $1-e^{-w}$ proves \eqref{eq:ch1_base_hazard}. Also
\begin{equation}
 \log w=-\frac{\log(1+\gamma z)}{\gamma}\longrightarrow-z
 \quad\text{as }\gamma\longrightarrow0.
\end{equation}
Thus the Gumbel branch is the continuous limit of both survival and density. All hazards are nonnegative on their domains. The factor $1/t$ is the Jacobian of the log-time transformation and must be retained.

For $0<v<1$, solve $S_0(t)=v$ by defining $r_v=-\log(1-v)>0$. The inverse survival is
\begin{equation}\label{eq:baseline_inverse}
 S_0^{-1}(v)=\begin{cases}
 \exp\{\mu+\sigma(r_v^{-\gamma}-1)/\gamma\},&\gamma\ne0,\\
 \exp\{\mu-\sigma\log r_v\},&\gamma=0.
 \end{cases}
\end{equation}
For a latent survival value $s\in(0,1)$, $S_i(t)=s$ is therefore inverted as $t=S_0^{-1}(s^{1/a_i})$, with $a_i=\exp(\boldsymbol z_i^T\boldsymbol\beta)$. This proves both the covariate adjustment and the exponentiation required to simulate positive event times.

\subsection{Likelihood factorization and all four substitutions}\label{app:part3}
For an observed finite event, the joint contribution is $f_{p,i}(t_i)$ multiplied by the censoring survival. For a censored observation, it is $S_{p,i}(t_i)$ multiplied by the censoring density. Conditional independent censoring with distinct parameters permits the censoring factors to be omitted, proving \eqref{non-infoL}.

For a censored subject, the two joint masses are
\begin{equation}
 P(Y_i>t_i,J_i=0)=p_{0i},\qquad
 P(Y_i>t_i,J_i=1)=S_{p,i}(t_i)-p_{0i}.
\end{equation}
Taking logs proves \eqref{eq:status_loglik}; summing over $J_i\in\{0,1\}$ recovers the observed censored contribution. The susceptible censored contributions specialize to
\begin{align}
 S_p-p_0&=\frac{\theta s}{1+\theta}&&\text{(Bernoulli)},\\
 S_p-p_0&=e^{-\theta}(e^{\theta s}-1)&&\text{(Poisson)},\\
 S_p-p_0&=\frac{(1-\theta)\theta s}{1-\theta s}&&\text{(geometric)},\\
 S_p-p_0&=\frac{-\log(1-\theta s)-\theta s}{[-\log(1-\theta)]s}&&\text{(shifted logarithmic)}.
\end{align}
Together with the densities from Supplementary Appendix~\ref{app:part1}, these identities prove \eqref{eq:lc_Ber_nonspatial}--\eqref{eq:lc_l_nonspatial}. They also show why the geometric and logarithmic censored terms must subtract the cure mass. For the latter, $-\log(1-x)-x=\sum_{k=2}^{\infty}x^k/k>0$. Functions such as \texttt{log1p} and \texttt{expm1}, or limiting series, are needed to avoid cancellation near zero.

This expression does not impute censored event times. To display all four models compactly, set $s_i=S_i(t_i)$, $d_i=f_i(t_i)$ and $\theta=\theta_i$ within each summand. Substitution gives
\begin{align}
 \ell_J^{\rm Ber}&=\sum_{i\in\mathcal E}\log\frac{\theta_i d_i}{1+\theta_i}
 +\sum_{i\in\mathcal C}\{J_i\log(\theta_i s_i)-\log(1+\theta_i)\},\label{eq:lc_Ber_nonspatial}\\
 \ell_J^{\rm Poi}&=\sum_{i\in\mathcal E}\{\log\theta_i+\log d_i-\theta_i(1-s_i)\}\notag\\
 &\quad+\sum_{i\in\mathcal C}\{-\theta_i+J_i\log[\exp(\theta_i s_i)-1]\},\label{eq:lc_p_nonspatial}\\
 \ell_J^{\rm Geo}&=\sum_{i\in\mathcal E}\{\log\theta_i+\log(1-\theta_i)+\log d_i-2\log(1-\theta_i s_i)\}\notag\\
 &\quad+\sum_{i\in\mathcal C}\{\log(1-\theta_i)+J_i\log[\theta_i s_i/(1-\theta_i s_i)]\}.\label{eq:lc_g_nonspatial}
\end{align}
For the shifted logarithmic model define $L_i=-\log(1-\theta_i)>0$, $A_i=-\log(1-\theta_i s_i)$ and
\begin{equation}\label{eq:lc_l_nonspatial}
\begin{split}
 D_i&=\frac{d_i}{L_i s_i^2}\left\{\frac{\theta_i s_i}{1-\theta_i s_i}-A_i\right\},\qquad
 B_i=\frac{A_i-\theta_i s_i}{L_i s_i},\\
 \ell_J^{\rm Log}&=\sum_{i\in\mathcal E}\log D_i+
 \sum_{i\in\mathcal C}\{(1-J_i)\log(\theta_i/L_i)+J_i\log B_i\}.
\end{split}
\end{equation}
Here $D_i=f_{p,i}(t_i)$ and $B_i=S_{p,i}(t_i)-p_{0i}$. Products with a zero indicator are omitted, rather than numerically evaluating $0\log0$. At $s_i=0$, expressions are interpreted through the original probabilities and their limits. In particular, an individual censored beyond a finite upper endpoint must have $J_i=0$.

When finite event times are augmented, a susceptible subject instead contributes the joint density $f_{p,i}(y_i)$, and a cured subject contributes $p_{0i}$. Integrating the former over $(t_i,\infty)$ gives $S_{p,i}(t_i)-p_{0i}$. This proves that \eqref{eq:full_aug_loglik} marginalizes to \eqref{eq:status_loglik} and then to \eqref{non-infoL}. No extra susceptibility factor should be multiplied into $f_{p,i}$.

\subsection{Conditional sampling and exact inversions}\label{app:part4}
\begin{proposition}[Conditional augmentation]\label{prop:conditional}
If $S_{p,i}(c)>0$, the conditional susceptibility probability is $(S_{p,i}(c)-p_{0i})/S_{p,i}(c)$. If the numerator is positive, the conditional finite event law given susceptibility and $Y_i>c$ is \eqref{eq:ch1:sem_step3}--\eqref{eq:conditional_cdf}.
\end{proposition}
\begin{proof}
Bayes' rule applied to the joint masses in Supplementary Appendix~\ref{app:part3} gives the susceptibility probability. Integrating the density $f_{p,i}$ over $(c,y]$ gives $S_{p,i}(c)-S_{p,i}(y)$; dividing by its total mass $S_{p,i}(c)-p_{0i}$ yields a CDF increasing from zero to one. Setting it equal to $U\sim\mathrm{Uniform}(0,1)$ gives \eqref{eq:conditional_inverse}. If instead $\widetilde H=(S_{p,i}(c)-S_{p,i}(y))/S_{p,i}(c)$, then $\widetilde H=q_i H$. Consequently, drawing $V\sim\mathrm{Uniform}(0,q_i)$ and solving $\widetilde H(y)=V$ is exactly equivalent. A draw on $(0,1)$ with this defective CDF would not be equivalent.
\end{proof}

For a target $B\in(p_{0i},1)$, inversion of $S_{p,i}(y)=B$ first yields $s=S_i(y)$:
\begin{equation}\label{eq:population_inverse}
 s=\begin{cases}
 [(1+\theta_i)B-1]/\theta_i,&\text{Bernoulli},\\
 1+\log B/\theta_i,&\text{Poisson},\\
 [1-(1-\theta_i)/B]/\theta_i,&\text{geometric},\\
 G_{\rm Log}^{-1}\{B G_{\rm Log}(\theta_i)\}/\theta_i,&\text{shifted logarithmic}.
 \end{cases}
\end{equation}
The last inverse is a unique root on $(0,\theta_i)$ because $G'_{\rm Log}>0$. Then use \eqref{eq:baseline_inverse} with $v=s^{1/a_i}$. For conditional sampling, take $B$ from \eqref{eq:conditional_inverse}. For unconditional simulation, draw $V\sim\mathrm{Uniform}(0,1)$, set $Y_i=\infty$ if $V\leq p_{0i}$, and otherwise solve $S_{p,i}(Y_i)=V$. This reproduces both the cure mass and the finite-time distribution. Equivalently, sample $M_i$, sample $M_i$ independent latent lifetimes using Supplementary Appendix~\ref{app:part2}, and take their minimum.

\subsection{An explicit EM comparator and the SEM distinction}\label{app:part5}
For a status-only augmentation, $E[J_i\mid\mathcal O;\boldsymbol\xi^{(r)}]=q_i^{(r)}$ for censored subjects. The exact EM objective is therefore
\begin{equation}\label{eq:em_q}
\begin{split}
 Q_J(\boldsymbol\xi\mid\boldsymbol\xi^{(r)})
 &=\sum_{i\in\mathcal E}\log f_{p,i}(t_i;\boldsymbol\xi)\\
 &\quad+\sum_{i\in\mathcal C}\big\{(1-q_i^{(r)})\log p_{0i}(\boldsymbol\xi)
       +q_i^{(r)}\log[S_{p,i}(t_i;\boldsymbol\xi)-p_{0i}(\boldsymbol\xi)]\big\}.
\end{split}
\end{equation}
The E-step is closed form for all four models; numerical M-step optimization is still required. Thus intractability of the E-step is not, by itself, an established advantage of SEM in this particular formulation. Replacing $q_i^{(r)}$ by a Bernoulli draw gives a status-only SEM. If event times are also drawn, the correct objective is $\ell_A$, not $\ell_J$.

\begin{proposition}[Conditional mean of the stochastic objective]
For a fixed candidate $\boldsymbol\xi$ for which the expectations exist, the sampled log-likelihood has conditional expectation equal to the EM objective associated with the same augmentation.
\end{proposition}
\begin{proof}
For status-only augmentation this follows by linearity in $J_i$. For status-and-time augmentation, a censored contribution has conditional expectation
\begin{equation}
 (1-q_i^{(r)})\log p_{0i}(\boldsymbol\xi)
 +q_i^{(r)}\int_{t_i}^{\infty}g_i(y;\boldsymbol\xi^{(r)})\log f_{p,i}(y;\boldsymbol\xi)\,dy.
\end{equation}
Summing gives $Q_A$. This identity concerns objective values, not maximizers: in general $E[\operatorname{argmax}\ell_A]\ne\operatorname{argmax}E[\ell_A]$. It does not prove that the finite-run SEM average is an exact MLE.
\end{proof}
For exact EM, writing $k_{\boldsymbol\xi}$ for the conditional augmentation law gives, when the terms are finite and the relevant support condition holds,
\begin{equation}
 \ell_o(\boldsymbol\xi)-\ell_o(\boldsymbol\xi^{(r)})
 =Q(\boldsymbol\xi\mid\boldsymbol\xi^{(r)})-Q(\boldsymbol\xi^{(r)}\mid\boldsymbol\xi^{(r)})
 +\mathrm{KL}(k_{\boldsymbol\xi^{(r)}}\Vert k_{\boldsymbol\xi}).
\end{equation}
Nonnegativity of the KL divergence proves monotonicity for an exact or generalized EM step increasing $Q$. Maximizing one random completion does not ensure that $Q$ increases, so this proof does not establish monotonicity for SEM.

\subsection{Score, information and uncertainty in cure probabilities}\label{app:part6}
Writing $s_i=S_i(t_i)$, $d_i=f_i(t_i)$ and $x_i^*=\theta_i s_i$, \eqref{eq:observed_loglik} becomes
\begin{equation}\label{eq:compact_observed}
 \ell_o=\sum_i\left[\delta_i\{\log\theta_i+\log d_i+\log G'(x_i^*)\}
 +(1-\delta_i)\log G(x_i^*)-\log G(\theta_i)\right].
\end{equation}
All derivatives below are local, within a region where the relevant probabilities are positive and differentiable. For any parameter component $v$,
\begin{align}\label{eq:score_general}
 \partial_v\ell_o=\sum_i\bigg[&\delta_i\left\{\frac{\partial_v\theta_i}{\theta_i}
 +\partial_v\log d_i+\frac{G''(x_i^*)}{G'(x_i^*)}\partial_v x_i^*\right\}\notag\\
 &+(1-\delta_i)\frac{G'(x_i^*)}{G(x_i^*)}\partial_v x_i^*
 -\frac{G'(\theta_i)}{G(\theta_i)}\partial_v\theta_i\bigg],
\end{align}
where $\partial_v x_i^*=s_i\partial_v\theta_i+\theta_i\partial_v s_i$. For example,
\begin{equation}
 \partial_{\beta_j}s_i=s_i a_i z_{ij}\log S_0(t_i),\qquad
 \partial_{\beta_j}\log d_i=z_{ij}[1+a_i\log S_0(t_i)].
\end{equation}
For incidence coefficients, $\partial_{b_j}\theta_i=\dot\theta_i x_{ij}$, where $\dot\theta_i=\theta_i$ for the exponential link and $\theta_i(1-\theta_i)$ for the logistic link.

For direct differentiation of the baseline, put $w=h^{-1/\gamma}$ and $r_v=\partial_v\log w$. For $\gamma\ne0$,
\begin{equation}
 r_\mu=\frac1{\sigma h},\qquad r_\sigma=\frac{z}{\sigma h},\qquad
 r_\gamma=\frac{\log h}{\gamma^2}-\frac{z}{\gamma h}.
\end{equation}
The accompanying derivatives of $\log h$ are $-\gamma/(\sigma h)$, $-\gamma z/(\sigma h)$ and $z/h$, respectively. Hence, for $v\in\{\mu,\sigma,\gamma\}$,
\begin{equation}
 \partial_v\log S_0=\frac{w}{e^w-1}r_v,\qquad
 \partial_v\log f_0=-\frac{\mathbf1\{v=\sigma\}}{\sigma}+(1-w)r_v-\partial_v\log h.
\end{equation}
At $\gamma=0$, use the continuous limits
\begin{equation}
 (r_\mu,r_\sigma,r_\gamma)=\left(\frac1\sigma,\frac z\sigma,\frac{z^2}{2}\right),\qquad
 (\partial_\mu\log h,\partial_\sigma\log h,\partial_\gamma\log h)=(0,0,z).
\end{equation}
These follow from $\log w=-z+\gamma z^2/2+O(\gamma^2)$ at fixed $t,\mu,\sigma$. Finally,
$\partial_v\log s_i=a_i\partial_v\log S_0$ and
$\partial_v\log d_i=\partial_v\log f_0+(a_i-1)\partial_v\log S_0$.

An explicit Hessian follows by defining $A(x)=\log G(x)$, $B(x)=\log G'(x)$ and $c_i=\log\theta_i+\log d_i$:
\begin{align}
 \partial_{uv}^2\ell_o=\sum_i\big[&\delta_i\{\partial_{uv}^2c_i+B''(x_i^*)\partial_u x_i^*\partial_v x_i^*
                  +B'(x_i^*)\partial_{uv}^2x_i^*\}\notag\\
 &+(1-\delta_i)\{A''(x_i^*)\partial_u x_i^*\partial_v x_i^*
                  +A'(x_i^*)\partial_{uv}^2x_i^*\}\notag\\
 &-A''(\theta_i)\partial_u\theta_i\partial_v\theta_i
 -A'(\theta_i)\partial_{uv}^2\theta_i\big],\label{eq:hessian_general}
\end{align}
where $A''=G''/G-(G'/G)^2$ and $B''=G'''/G'-(G''/G')^2$. This provides an independent target for checking numerical Hessians.

If differentiation under the augmentation integral is justified, let $U_c=\nabla\ell_c$ and $H_c=\nabla^2\ell_c$. Differentiating the marginal likelihood gives the score identity and observed-information decomposition
\begin{equation}\label{eq:missing_information}
 U_o=E(U_c\mid\mathcal O),\qquad
 \mathcal I_o=E(-H_c\mid\mathcal O)-\operatorname{Var}(U_c\mid\mathcal O).
\end{equation}
To see the second identity, differentiate the first, using $\nabla\log k_{\boldsymbol\xi}=U_c-U_o$; the extra term is the conditional covariance of $U_c$. Omitting this missing-information term is generally incorrect. Parameter-dependent integration limits require explicit justification; the identity is not an unconditional regularity claim for all GEV parameters.

Finally, with $\eta_i=\boldsymbol x_i^T\boldsymbol b$,
\begin{equation}\label{eq:cure_gradient}
 \frac{\partial p_{0i}}{\partial\eta_i}
 =-\frac{a_0G'(\theta_i)\dot\theta_i}{G(\theta_i)^2}<0.
\end{equation}
It equals $-p_{0i}(1-p_{0i})$ for Bernoulli and geometric models and $-\theta_i p_{0i}$ for Poisson. For the shifted logarithmic model, with $L=-\log(1-\theta_i)$, it is
\begin{equation}
 \theta_i(1-\theta_i)\frac{L-\theta_i/(1-\theta_i)}{L^2}<0.
\end{equation}
Thus a positive incidence coefficient decreases cure probability under every link used here. It is not a population log-hazard ratio. If $g_i=\nabla_{\boldsymbol\xi}p_{0i}$ and $\widehat V$ is valid, the delta method gives
\begin{equation}
 \widehat{\operatorname{Var}}(\widehat p_{0i})=g_i^T\widehat V g_i,\qquad
 \widehat{\operatorname{Var}}(\widehat p_{0i}-\widehat p_{0j})=(g_i-g_j)^T\widehat V(g_i-g_j).
\end{equation}
The covariance is essential when comparing predictions from the same fitted model. A logit-scale interval, using gradient $g_i/[p_{0i}(1-p_{0i})]$, respects the probability range.

\subsection{Endpoint behavior and a limitation of unrestricted maximization}\label{app:part7}
The following result concerns the observed-data likelihood in \eqref{non-infoL}, with the parameter-independent censoring factors omitted. It is a statement about the statistical model, not about the behavior of a particular optimizer.

\begin{proposition}[Unbounded likelihood on the unrestricted parameter space]\label{prop:unbounded}
Fix a dataset with finite covariates and $0<t_i<\infty$ for every subject. Suppose that $\mathcal E=\{i:\delta_i=1\}$ is nonempty. Assume that the parameter space contains a submodel with $\boldsymbol\beta=0$, some fixed $\sigma>0$, some fixed $\gamma=-g$ with $g>1$, fixed admissible incidence coefficients $\boldsymbol b$, and all location parameters in a neighborhood of
$\log y_*-\sigma/g$, where $y_*=\max_{i\in\mathcal E}t_i$.
Then the observed likelihood is unbounded above for each of the Bernoulli, Poisson, geometric and shifted logarithmic models. In particular, there is no finite-valued global maximum over the unrestricted parameter space.
\end{proposition}
\begin{proof}
Let $\mathcal E_* =\{i\in\mathcal E:t_i=y_*\}$ and $k_*=|\mathcal E_*|\geq1$. Fix all parameters except the location, and define the sequence
\begin{equation}\label{eq:unbounded_path}
 \mu_\varepsilon=\log y_*+\varepsilon-\frac{\sigma}{g},\qquad
 \gamma=-g,\qquad\boldsymbol\beta=0,\qquad\varepsilon\downarrow0.
\end{equation}
The finite upper endpoint and the support function are
\begin{equation}\label{eq:unbounded_support}
 t_U(\varepsilon)=\exp(\mu_\varepsilon+\sigma/g)=y_*e^\varepsilon,
 \qquad h_\varepsilon(t)=\frac{g}{\sigma}\left\{\log\frac{y_*}{t}+\varepsilon\right\}.
\end{equation}
Every observed event is strictly below $t_U(\varepsilon)$ for every $\varepsilon>0$. Thus this sequence stays within the event-support domain rather than assigning an observed event zero probability.

\emph{Contribution of events at the largest event time.}
At $t=y_*$, write $h_*=g\varepsilon/\sigma$. Equations~\eqref{eq:ch1_base_surv} and \eqref{eq:baseline_density} give
\begin{align}
 S_{0,\varepsilon}(y_*)&=1-e^{-h_*^{1/g}}\sim h_*^{1/g},\label{eq:unbounded_survival}\\
 f_{0,\varepsilon}(y_*)&=\frac{h_*^{1/g-1}e^{-h_*^{1/g}}}{\sigma y_*}
 \sim\frac{(g/\sigma)^{1/g-1}}{\sigma y_*}\varepsilon^{1/g-1}.\label{eq:unbounded_density}
\end{align}
The symbol $u_\varepsilon\sim v_\varepsilon$ means $u_\varepsilon/v_\varepsilon\to1$. Because $g>1$, the exponent $1/g-1$ is negative. With $\boldsymbol\beta=0$, $a_i=1$, so the covariate-adjusted latent density equals $f_{0,\varepsilon}$. For the four generating functions,
\begin{equation}\label{eq:positive_first_coefficient}
 G'(0)=\begin{cases}
 1,&\text{Bernoulli, Poisson or geometric},\\
 1/2,&\text{shifted logarithmic}.
 \end{cases}
\end{equation}
Consequently, for every $i\in\mathcal E_*$,
\begin{equation}\label{eq:unbounded_population_density}
 f_{p,i,\varepsilon}(y_*)
 =\frac{\theta_iG'(\theta_iS_{0,\varepsilon}(y_*))}{G(\theta_i)}f_{0,\varepsilon}(y_*)
 \sim C_i\varepsilon^{1/g-1},
\end{equation}
where
\begin{equation}
 C_i=\frac{\theta_iG'(0)}{G(\theta_i)}\,
      \frac{(g/\sigma)^{1/g-1}}{\sigma y_*}\in(0,\infty).
\end{equation}
Thus all tied largest events have diverging density contributions, not just one selected event.

\emph{Contribution of earlier events.}
For $i\in\mathcal E\setminus\mathcal E_*$,
$h_\varepsilon(t_i)\to(g/\sigma)\log(y_*/t_i)>0$.
Hence $S_{0,\varepsilon}(t_i)$ converges to a value strictly between zero and one, and $f_{0,\varepsilon}(t_i)$ converges to a finite positive value. Since $\theta_i$ is fixed and admissible and $G'$ is finite and positive at the corresponding argument, $f_{p,i,\varepsilon}(t_i)\to d_i^*\in(0,\infty)$.

\emph{Contribution of censored observations.}
For every $i\in\mathcal C=\{i:\delta_i=0\}$ and every $\varepsilon>0$,
\begin{equation}\label{eq:censoring_lower_bound}
 S_{p,i,\varepsilon}(t_i)=\frac{G(\theta_iS_{0,\varepsilon}(t_i))}{G(\theta_i)}
 \geq\frac{G(0)}{G(\theta_i)}=p_{0i}>0.
\end{equation}
If $t_i<y_*$, this contribution converges to a finite positive survival probability. If $t_i=y_*$, it converges to $p_{0i}$. If $t_i>y_*$, then $t_i>t_U(\varepsilon)$ for all sufficiently small $\varepsilon$, and its contribution is exactly $p_{0i}$ under the support extension. In particular, censoring beyond the susceptible upper endpoint is possible because the cure mass is at infinity; such censoring does not invalidate the sequence.

\emph{Combining the contributions.}
All factors not belonging to $\mathcal E_*$ have finite positive limits. Empty products, if there are no earlier events or no censored subjects, equal one. Therefore, for a finite constant $C>0$,
\begin{equation}\label{eq:likelihood_divergence_rate}
 L_o(\boldsymbol\xi_\varepsilon)\sim C\varepsilon^{k_*(1/g-1)}\longrightarrow\infty,
\end{equation}
and equivalently
\begin{equation}\label{eq:loglik_divergence_rate}
 \ell_o(\boldsymbol\xi_\varepsilon)
 =k_*\left(1-\frac1g\right)\log(1/\varepsilon)+\log C+o(1)
 \longrightarrow\infty.
\end{equation}
For each fixed positive $\varepsilon$, every event is inside the support and all likelihood contributions are finite. The supremum over these feasible parameter vectors is nevertheless infinite. A parameter vector with a finite likelihood value therefore cannot be a global maximizer.
\end{proof}

\begin{proposition}[Endpoint order with fixed covariate effects]\label{prop:endpoint_covariates}
Fix $\gamma=-g<0$, $\sigma>0$, admissible incidence parameters and finite covariate effects $a_i=\exp(\boldsymbol z_i^T\boldsymbol\beta)>0$. For a fixed finite upper endpoint $t_U$, as $t\uparrow t_U$,
\begin{equation}\label{eq:covariate_endpoint_order}
 f_{p,i}(t)\sim
 \frac{\theta_iG'(0)}{G(\theta_i)}\frac{a_i}{g}
 \left(\frac{g}{\sigma t_U}\right)^{a_i/g}
 (t_U-t)^{a_i/g-1}.
\end{equation}
Along the location sequence \eqref{eq:unbounded_path}, with these fixed covariate effects instead of $\boldsymbol\beta=0$, the likelihood is unbounded whenever
\begin{equation}\label{eq:general_divergence_condition}
 \sum_{i\in\mathcal E_*}\left(\frac{a_i}{g}-1\right)<0.
\end{equation}
\end{proposition}
\begin{proof}
Let $\Delta=t_U-t\downarrow0$. Since
$h(t)=(g/\sigma)\log(t_U/t)\sim g\Delta/(\sigma t_U)$, put $c=g/(\sigma t_U)$ to obtain
\begin{equation}
 S_0(t)\sim(c\Delta)^{1/g},\qquad
 f_0(t)\sim\frac{c^{1/g}}{g}\Delta^{1/g-1}.
\end{equation}
Substituting into the exact identity $f_i=a_i f_0S_0^{a_i-1}$ gives
$f_i(t)\sim(a_i/g)c^{a_i/g}\Delta^{a_i/g-1}$.
Also $S_i=S_0^{a_i}\to0$, so $G'(\theta_iS_i)\to G'(0)>0$, proving \eqref{eq:covariate_endpoint_order}. Along \eqref{eq:unbounded_path}, a direct substitution at $y_*$ gives a finite positive constant $\widetilde C_i$ such that
$f_{p,i,\varepsilon}(y_*)\sim\widetilde C_i\varepsilon^{a_i/g-1}$.
Earlier event and censored contributions still have positive finite limits by the preceding proof. Multiplication gives
\begin{equation}
 L_o(\boldsymbol\xi_\varepsilon)\sim\widetilde C\,
 \varepsilon^{\sum_{i\in\mathcal E_*}(a_i/g-1)},\qquad\widetilde C>0,
\end{equation}
which diverges under \eqref{eq:general_divergence_condition}.
\end{proof}
For a unique largest event, the sufficient condition is simply $a_i<|\gamma|$ for that event. With $a_i=1$ for all subjects, this reduces to $\gamma<-1$, recovering Proposition~\ref{prop:unbounded}. Conversely, failure of \eqref{eq:general_divergence_condition} only removes divergence along this particular path; it does not establish boundedness of the likelihood over the entire parameter space.

These are sufficient-condition results, not a complete classification of all possible likelihood singularities. They concern the stated unrestricted model and do not show that an optimizer using additional bounds or penalties must diverge. Nor do they prove identifiability, estimator consistency, asymptotic normality or SEM ergodicity.

\subsection{Existence on a constrained parameter set}\label{app:compact}
\begin{proposition}[A sufficient constrained existence condition]\label{prop:compact}
Let $K$ be a nonempty compact parameter set with $\sigma\geq\sigma_{\min}>0$, admissible incidence parameters, and, for a fixed $\varepsilon>0$,
\begin{equation}\label{eq:margin}
 1+\gamma\frac{\log t_i-\mu}{\sigma}\geq\varepsilon\qquad\text{for every observed event }i\in\mathcal E.
\end{equation}
Use the continuous $\gamma=0$ branch and the support extensions for censored observations. Then $\ell_o$ attains a finite maximum on $K$.
\end{proposition}
\begin{proof}
On $K$, bounded regression coefficients keep $\theta_i$ within a compact subset of its admissible domain and $a_i$ positive and bounded. The event margin and positive scale bound make each event density a finite, positive continuous function, including at $\gamma=0$. The extended baseline survival is continuous at support boundaries. Each censored population survival is continuous and at least $p_{0i}$, whose minimum on $K$ is positive. Thus the finite sum of log contributions $\ell_o$ is continuous on the nonempty compact set $K$. The extreme value theorem gives a finite attained maximum.
\end{proof}
This is a sufficient condition for a \emph{constrained} estimator, not a claim that the original computations used $K$. It neither establishes uniqueness nor validates Wald intervals or SEM convergence. Bounds and the endpoint margin must be specified, justified and subjected to sensitivity analysis; sample-dependent restrictions also affect inference. For an augmented M-step, the support condition must additionally cover every imputed susceptible event. No numerical bounds are imposed retroactively in this draft.

\clearpage
\setcounter{table}{0}
\renewcommand{\thetable}{\Alph{section}\arabic{table}}
\renewcommand{\theHtable}{\Alph{section}\arabic{table}}

\section{Supplementary numerical example}\label{app:poisson_example}
Table~\ref{tab:poisson_sim_example} presents a selected result from the simulation study in Section~\ref{ch2:sim}: the Poisson generating and fitting model with a Type I baseline. The parameter vector is $(b_0,b_1,\beta_1,\mu,\sigma,\gamma)=(-0.7,0.52,0.64,0.2,0.17,0)$, with $n=1000$ and a reported censoring proportion of $0.743$. The study also considers the other latent-cause distributions and baseline settings described in the main text; this table is one selected numerical summary, not the full set of simulation results.

\begin{table}[H]
\centering
\caption{Selected Poisson simulation results with a Type I baseline and $n=1000$.}
\label{tab:poisson_sim_example}
\renewcommand{\arraystretch}{1.12}
\begin{tabular*}{\textwidth}{@{\extracolsep{\fill}}lrrrr@{}}
\toprule
Parameter & True value & Mean estimate & Bias & RMSE \\
\midrule
$b_0$ & $-0.700$ & $-0.651$ & $0.049$ & $0.069$ \\
$b_1$ & $0.520$ & $0.536$ & $0.016$ & $0.042$ \\
$\beta_1$ & $0.640$ & $0.642$ & $0.002$ & $0.010$ \\
$\mu$ & $0.200$ & $0.162$ & $-0.038$ & $0.050$ \\
$\sigma$ & $0.170$ & $0.152$ & $-0.018$ & $0.042$ \\
$\gamma$ & $0.000$ & $-0.032$ & $-0.032$ & $0.033$ \\
\bottomrule
\end{tabular*}
\par\smallskip
\begin{minipage}{\textwidth}\footnotesize
RMSE: root mean squared error. The generating and fitted latent-cause distributions are both Poisson.
\end{minipage}
\end{table}

\end{document}